# Vishap stelae as cult dedicated prehistoric monuments of Armenian Highlands: data analysis and interpretation

Vahe Gurzadyan[1,*], Arsen Bobokhyan[2]

[1] Center for Cosmology and Astrophysics, Alikhanian National Laboratory and Yerevan State University, Yerevan, Armenia

[2] Institute of Archaeology and Ethnography, National Academy of Sciences and Yerevan State University, Yerevan, Armenia

* Author for correspondence    gurzadyan@yerphi.am

Vishaps, or dragon stones, are prehistoric stelae discovered in the high-altitude mountainous regions of modern-day Armenia and adjacent regions. The first statistical analysis of their elevation distribution and size reveals that their construction was intentionally labor-intensive rather than arbitrary. The findings support the hypothesis that vishaps were closely associated with an ancient water cult, as they are predominantly situated near water sources, including high-altitude springs and discovered prehistoric irrigation systems. Furthermore, the unexpected bimodal distribution of their altitudes suggests specific placement patterns, potentially linked to seasonal human activities or ritual practices. These findings contribute to a deeper understanding of the symbolic and functional significance of vishap stelae within the framework of prehistoric social and religious systems.

## Introduction

Vishaps (from the Armenian word for "dragon") are stone stelae adorned with animal imagery, located in the high-altitude summer pastures of contemporary Armenia and neighboring regions, predominantly at elevations ranging from approximately 1000 to 3000 meters above sea level (see Figure 1). The vernacular term "vishap" is etymologically linked to local folklore.

Based on their form and iconography, the vishaps are classified into three primary typological categories, ranging from 110 to 550 cm in height, and crafted from locally available stone materials, primarily andesite and basalt. The first class, termed *piscis*, consists of stones carved and polished into a fish shape. The second category, *vellus*, includes stones shaped to resemble a stretched or draped bovid hide. The third class, the *hybrida*, merges the iconographies of the *piscis* and *vellus* types (see Figure 2–3).

The majority of vishaps are either collapsed or placed horizontally on the ground. However, all three typological groups of vishaps exhibit carving and polishing on all faces, with the "tail" invariably left uncarved. This consistent feature strongly suggests that vishaps were originally positioned upright.

The scholarly interest in vishaps emerged in the early twentieth century and was revisited sporadically afterwards[1-5]. Among these scholars, Ashkharbek Kalantar stands out for his unique contribution, who was the first to approach the vishaps within an archaeological context, linking them to other megalithic phenomena. Following a survey of vishap sites on Mount Aragats and their correlation with the surrounding cultural landscapes, Kalantar proposed the influential hypothesis that vishaps demarcate critical points in prehistoric irrigation systems (Kalantar[6,7,8], republished in English in [9-12]).

After Kalantar's research, the study of vishaps remained limited. To tackle the problem with advanced archaeological methods, in 2012, a collaboration started between the Institute of Archaeology and Ethnography, National Academy of Sciences of Armenia, the Free University of Berlin, and later also the Ca' Foscari University of Venice. The general aim of the collaboration was to investigate the function and socio-economic background of the vishap phenomenon with the toolbox of contemporary archaeological science, most importantly landscape survey and the stratigraphic excavation of selected contexts. During the research the systematic assessment of the key landscape features associated with vishaps was defined, which was followed by excavations at the site Tirinkatar (also known as Karmir Sar), discovered by the expedition on the southern slopes of Mount Aragats (Figure 4–5).

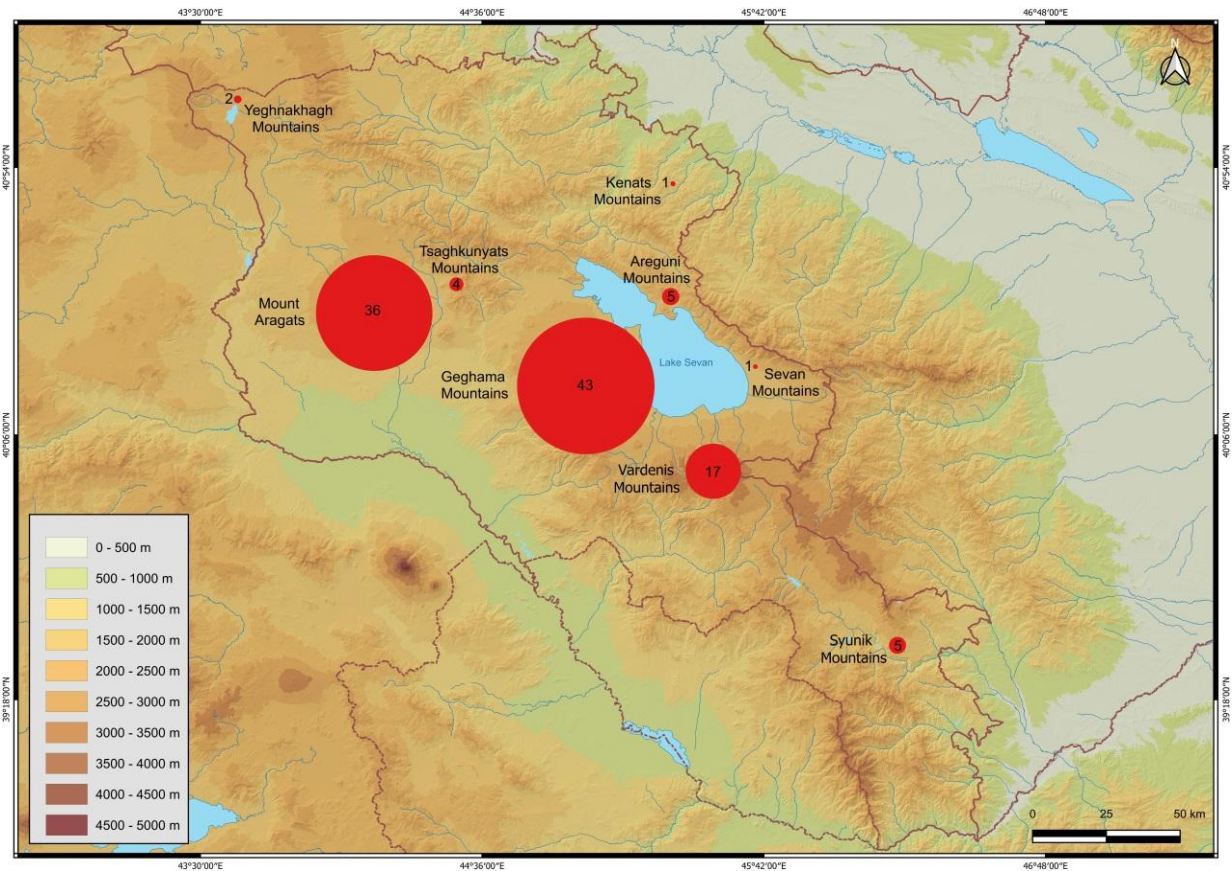

*Figure 1. Distribution and quantification of vishaps ("Vishap" Project).*

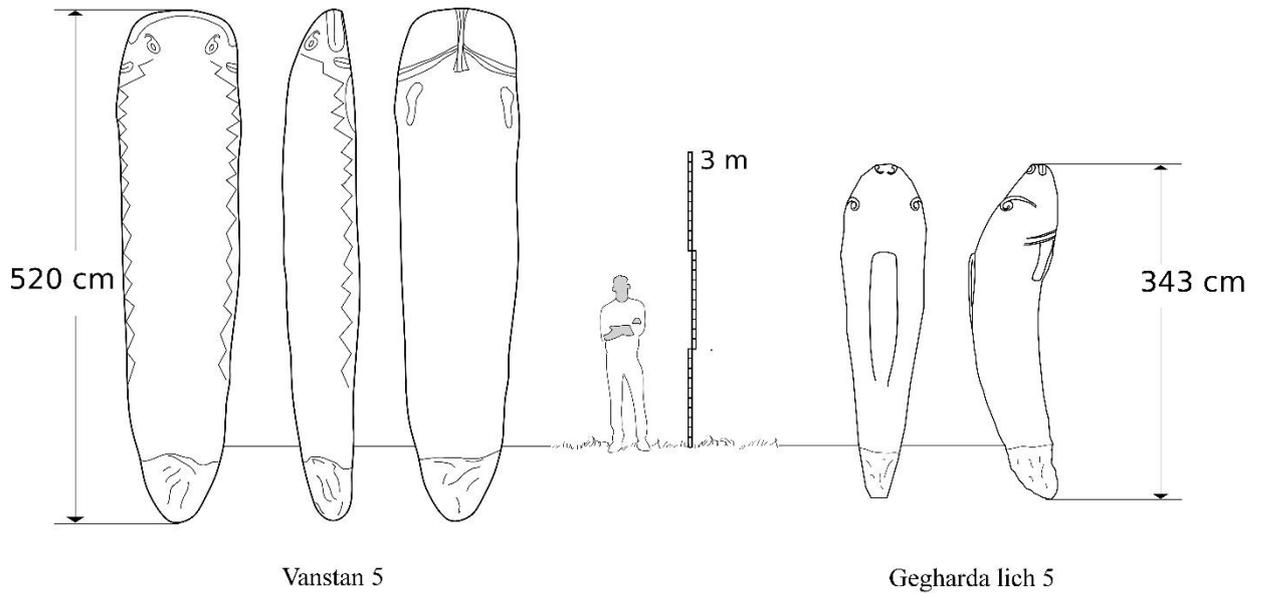
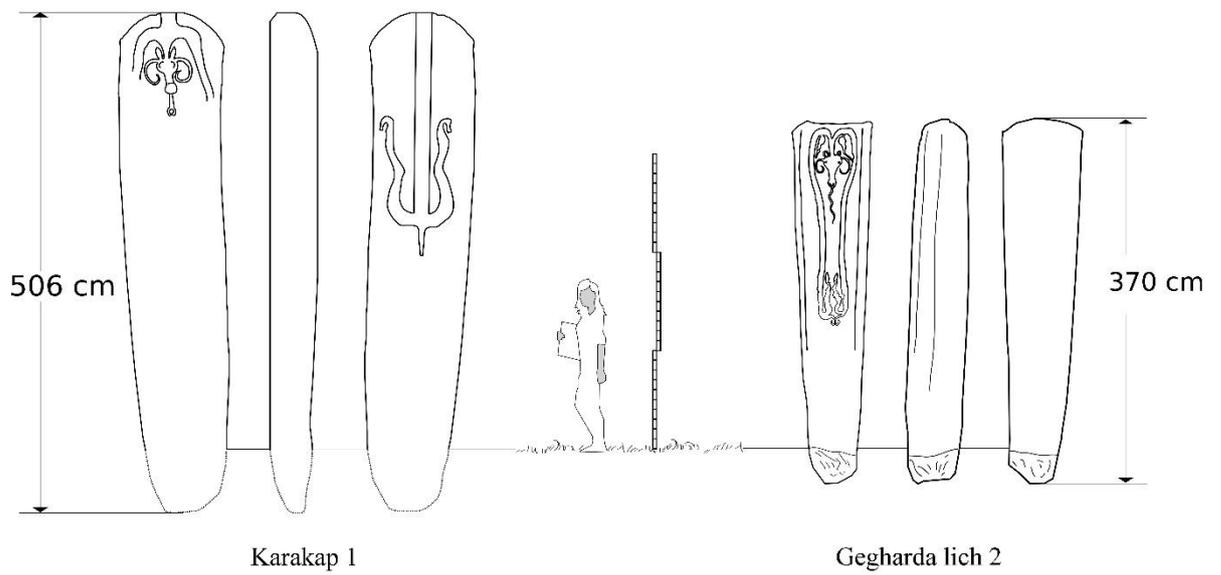
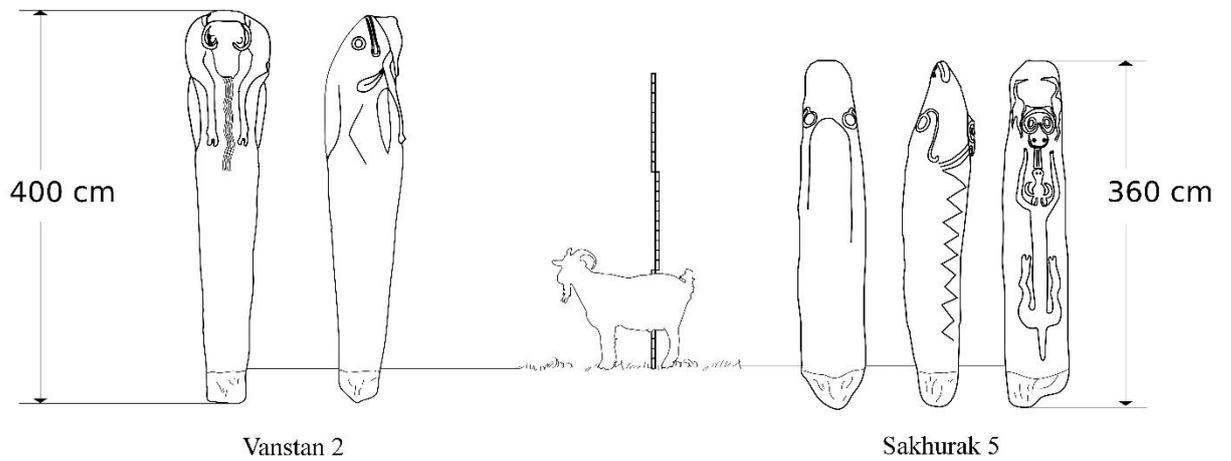

*Figure 2. Typology of vishaps: piscis, vellus, hybrida ("Vishap" Project, A. Gilibert).*

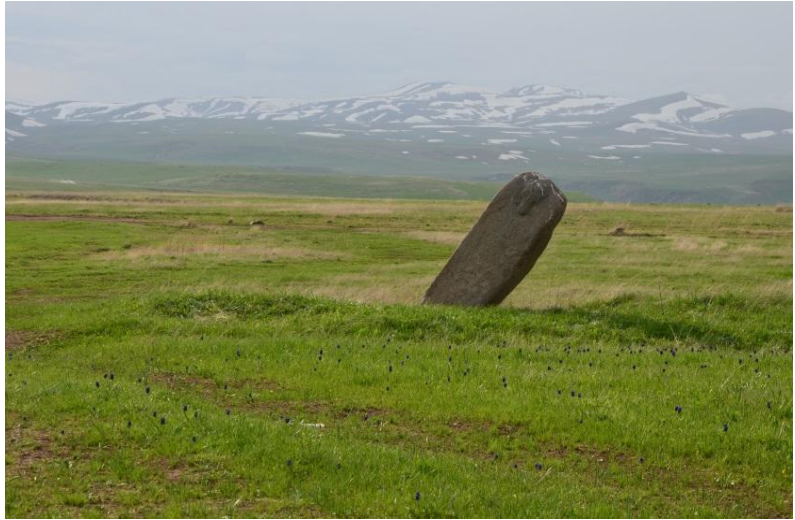

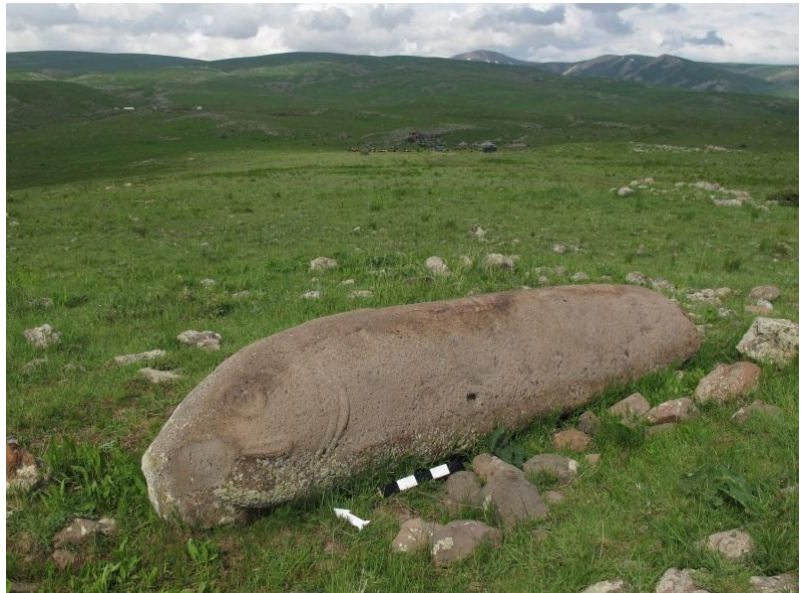
B

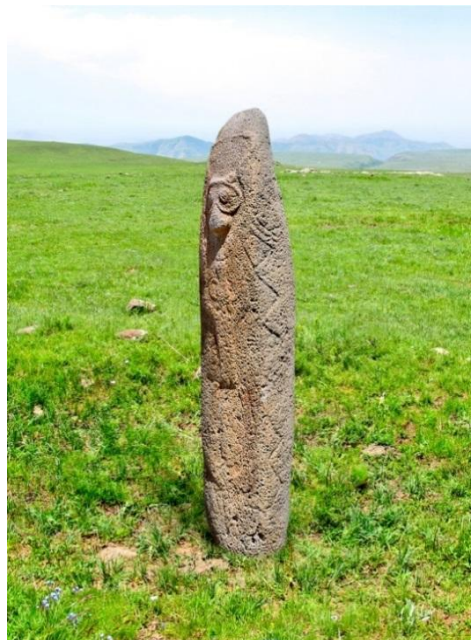
C

*Figure 3. Examples of A) vellus (Madinayi Sar 1), B) piscis (Vanstan 2) and C) hybrida (Sakhurak 5) vishaps, ("Vishap" Project', A. Bobobkhyan, reconstruction of the hybrida in its original setting by V. Mkrtchyan).*

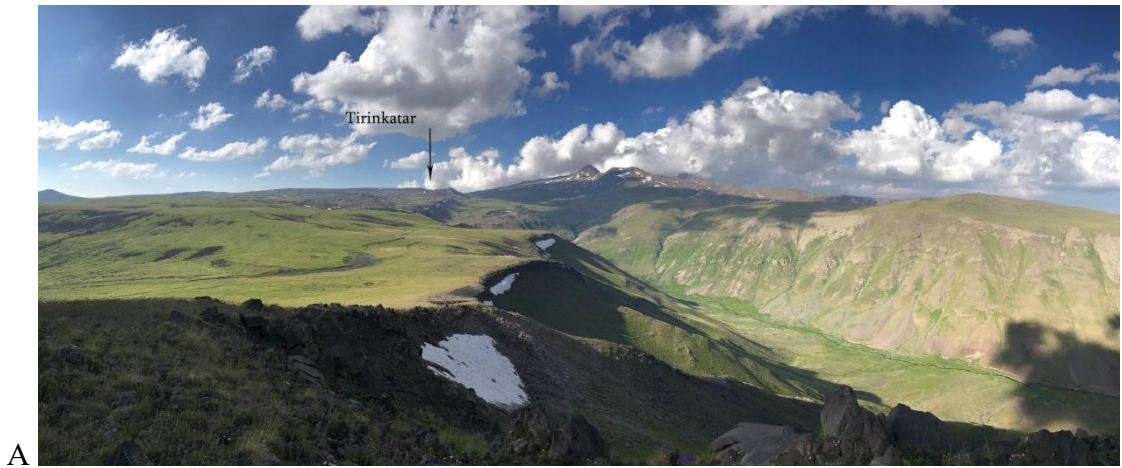

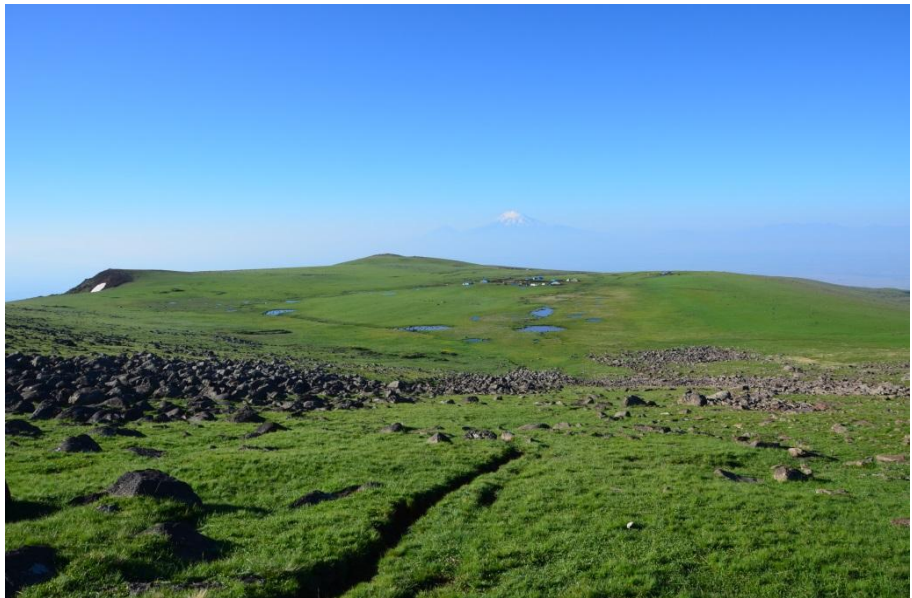

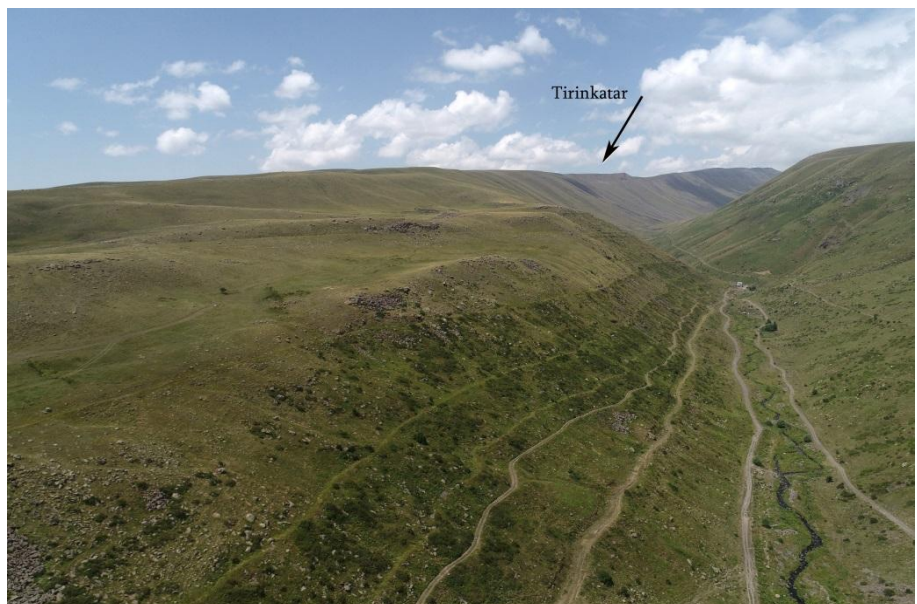

*Figure 4. A) Location of Tirinkatar on Mount Aragats; B) The sacred landscape of Tirinkatar, with concentration of 12 vishaps; C) Irrigation system "12 Canals" directly below Tirinkatar: it was used until the mid twentieth century, hence its precise dating is still questionable ("Vishap" Project, H. von der Osten, A. Hakhverdyan).*

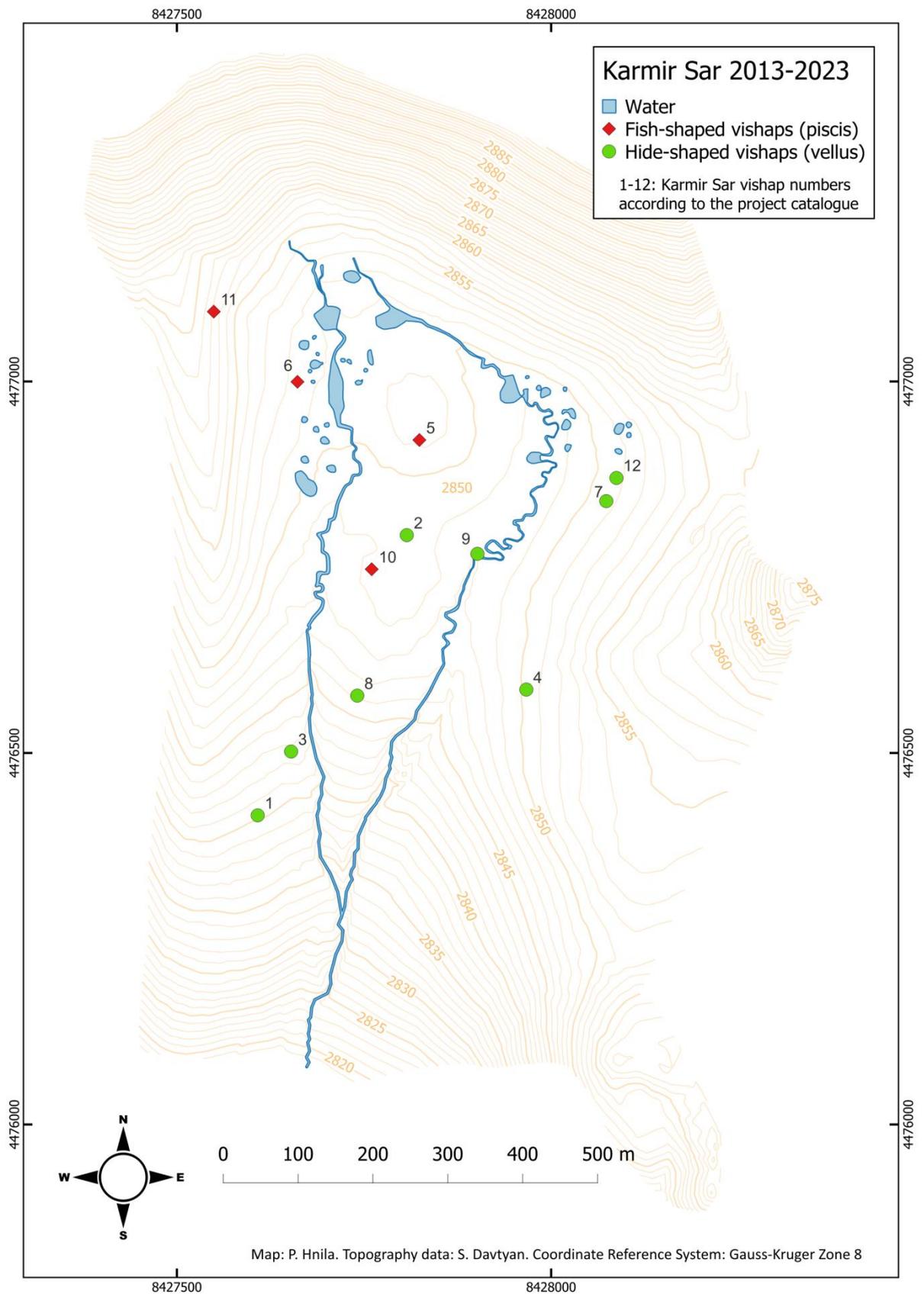

*Figure 5. Topographic map of Tirinkatar with location of vishaps; piscis vishaps are marked red, while vellus vishaps are marked green (Map: P. Hnila, Topography data: S. Davtyan).*

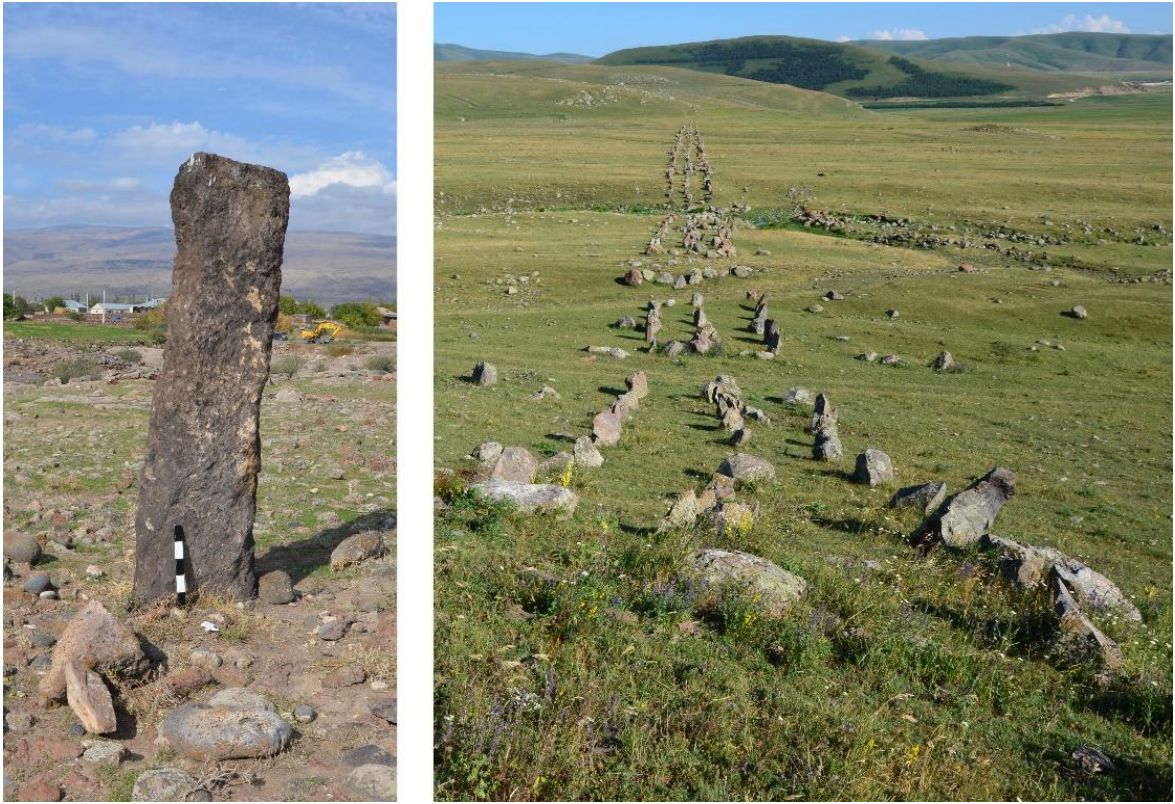

*Figure 6. A) Example of a menhir in Shamiram (Aragatsotn Region) and B) A menhir alignment in Hartashen (Shirak Region), Armenia, ca. mid second millennium BC ("Vishap" Project, A. Bobokhyan).*

Within the territory of modern Armenia, it was possible to collect information on all currently known vishaps, more than half of which were previously unpublished (Table 1). Some of them have been moved to different places. Even in most of these latter cases, the original location could be identified. Sometimes, vishaps were found not exactly in situ but still in loco, i.e., a few meters away from their supposed original setup location. These collected data allow us to study distribution patterns, function, and the dating of vishaps in a way that was not possible before.

We now know that vishaps were originally standing monuments, mostly associated with artificial platforms of medium-sized, unworked stones. In their immediate surroundings, cairns, cromlechs, tombs, petroglyphs, prehistoric settlements, and campsites frequently occur. Most landscapes typical of vishaps are situated in well-defined meadows, specifically at places where the landscape drops into slight depressions. These secluded meadows are typically ancient satellite volcanic craters, which are rich in water, sometimes even are marshy, and their concave form significantly reduces their overall visibility in the wider area [13-24].

A previous quantitative analysis of the correlation between vishap typology, size, and location based on a corpus of all vishaps [25] (p. 529-530) indicated that: a) Typology and size did not correlate significantly, with all three types of vishaps existing in varying sizes; b) The mean altitude of *piscis* vishaps was considerably higher than the mean altitude of *vellus* vishaps (2796 vs. 2493 meters a.s.l.), with a larger slice of the *piscis* vishaps located in the altitude range of 2700-3000 meters a.s.l. than the *vellus* vishaps (75% vs. 43%). Also, 15% of *vellus* vishaps are located below 2000 meters a.s.l., whereas not a single *piscis* vishap is located at or below that altitude.

This preliminary quantitative review suggests a specific correlation of the fish imagery with high-altitude locations, reinforcing our understanding that vishaps in general, and *piscis* vishaps in particular, may be connected to a water cult.

In the present study, we set forth a combined statistical analysis of two physical characteristics of the vishaps, bypassing the aspect of iconography: a) The vishaps' sizes, which we propose to correlate as a quantitative measure of the work effort required for their creation; b) The elevations of the vishaps' locations, as a quantitative indicator of the snow-free period available for human activity in mountains.

**Methods**

Periodization and dating

Situated in the northern part of the Fertile Crescent, the Armenian Highlands are recognized as one of the regions where the agro-pastoral way of life was established during the Neolithic (ca. 10000–5500/5200 BC) and Chalcolithic (ca. 5200–3500 BC) periods. During the Early Bronze Age (ca. 3500–2400 BC), the region was characterized by the Kura-Araxes cultural assemblage. The subsequent Middle Bronze Age (ca. 2400–1600 BC) saw the emergence of various sequential traditions, including the Kurgan, Trialeti-Vanadzor, Sevan-Artsakh, Karmir-Berd, and Karmir-Vank traditions, all of which exhibited economies largely centered on pastoralism. The Late Bronze (ca. 1600–1200 BC) and Early Iron (ca. 1200–900 BC) Age societies, associated with the "Lchashen-Metsamor" tradition, display clear indicators of complex agricultural economies and early state formation. These developments culminated in the establishment of the Urartian state, an imperial entity that utilized cuneiform script during the Middle Iron Age (ca. 900–600 BC), for details, see [26-28].

To which of the mentioned periods and traditions could the vishap belong?

The history of the problem of dating vishaps can be divided into certain stages. In the initial stage, the vishaps were considered an "archaic" phenomenon. After the discovery in 1963 of the vishap

Garni 1, with a secondary Urartian cuneiform inscription of King Argishti I, dating to the first half of the eighth century BC, the vishaps were definitely shifted into the second millennium BC. The next stage is connected with the stratigraphic investigations led by the "Vishap Project" at the archaeological site of Tirinkatar. In the course of several excavation operations, a total of 46 organic samples have been collected and radiocarbon dated. Archaeological data collected from excavated contexts indicate that, despite some interruptions, Tirinkatar was visited and used for campsite activities at least from the Neolithic period, i.e., from the end of the sixth millennium BC onwards. In two cases, samples from stratified contexts allow us to radiocarbon-date the erection of two vishaps around 4200–4000 BC. While these dates cannot automatically be extended to all vishaps across the Armenian Highlands, it is at least probable that vishaps are a Chalcolithic production, later reused and reembedded in many different ways. It means, at least since the Late Chalcolithic period, around the end of the fifth millennium BC, the site was used as a cultural landscape with an extraordinary concentrartion of vishaps. During the mid-3rd to the end of the second millennia BC, other archaeological features were added, including large aggregated cell structures, cromlechs, cairns, barrows, and petroglyphs. These data establish Tirinkatar as a key site that can be considered a model for landscape use and dating of vishaps, cf. [16,17]; for further discussions on dating cf. [29, 30].

Although certain chronological anchors efficient for dating in the ancient Near East [31-35] are still not directly applicable to the data at our disposal, the efforts to involve various laboratory methods must continue.

## Distribution patterns

The interpretation of the phenomenon of vishaps necessitates a detailed examination of their spatial distribution and positioning. This topic has been extensively explored within the context of European megalithic cultures (see, e.g. [30,36]). A pertinent question arises: can the findings from European megalithic monuments be applied to the analysis of vishaps? However, before addressing this, it is essential to determine whether vishaps can be appropriately compared to European megalithic objects.

In the 1920s, during the active efforts of the Committee for the Preservation of Antiquities of Armenia, vishaps were officially categorized as a form of megalithic monument[6,37]. Subsequent research included vishaps within the broader system of megalithic cultures based on certain formal resemblances. However, the polished, geometrically defined figures and iconography of vishaps clearly distinguish them as statues rather than menhirs in the European sense. Notably, examples of menhirs, resembling those found in Europe, are known to exist in Armenia, both as solitary standing stones and alignments (Figure 6).

Vishaps have been discovered across three distinct ecological zones, each defined by settlement patterns and agricultural activity: 1) The foothills (ca. 1000–2200 meters a.s.l.), which are characterized by permanent settlements; 2) Transitional zones between the foothills and highlands (ca. 2200–2400 meters a.s.l.), with few permanent settlements; and 3) The high-altitude highlands (ca. 2400–3000 meters a.s.l.), where permanent settlements are absent.

Vishaps have been found either as isolated, standalone monuments or in groups. Grouped vishaps are more common in transitional zones (e.g., Dashtademi Sar 1–2, Vanstan 1–5) and particularly in the highlands (e.g., Al Lake 1–3, Arshaluysi Sar 1–5, Gegharda Lich 1–8, Geghasar 1–3, Karakap 1–5, Lcharot 1–3, Sakhurak 1–6, Tirinkatar 1–12, Ughtuakunk 1–3). In groupings, a specific quantitative distribution is evident: significant clusters of vishaps are found within restricted areas and in substantial numbers, such as in Tirinkatar (12 examples), Gegharda Lich (8), Karakap (5), and Sakhurak (5) (cf. also Arshaluysi Sar – 5, Vanstan – 5).

It is noteworthy that the vishaps located in the foothills are not uniformly confined to compact areas but are dispersed across considerable distances from one another (e.g., Aghavnadzori Sar 1–4, Artanish 1–5, Garni 1–2, Harzhis 1–2, Lchashen 1–2, Taratumb 1–3), though they maintain a linear connection between sites.

Concerning the orientation of the vishaps, it is plausible that their spatial arrangement follows specific patterns. Like other monuments across the Ancient World (e.g., [36,38,39]), the reconstruction of the original orientation of vishaps remains challenging due to their frequent discovery in loco. However, given their considerable weight, it is improbable that these monuments were displaced over long distances.

The site of Tirinkatar exemplifies the difficulty in determining the original, precise orientation of standing vishaps, as most have been found in secondary contexts. However, in three specific cases (Tirinkatar 2, 8, and 10, see Figure 5), a tentative reconstruction of the original orientation was possible, based on the positions of the original pits and the direction in which the vishaps had fallen. In each of these cases, the orientation was toward the NNW, possibly directed at the peak of Mount Aragats or the site's primary water source (two of them published in [15], Figures 8 and 11).

An additional notable spatial pattern is the bifurcated distribution of vishap types at Tirinkatar: the *piscis* forms are generally situated in the northwestern portion of the site, near water accumulations, while the *vellus* vishaps are located in the northeastern and southern areas of the site (Figure 5).

## Materials and maps

This study focuses on vishap stelae with a specific emphasis on their dimensions and altitudinal distribution. Corresponding data were gathered through systematic field surveys. The dataset includes 115 vishaps documented in the Republic of Armenia (Table 1).

Table 1. List of the vishaps in the territory of the Republic of Armenia: their lengths and altitudes.

| ID, sequence number | Monument, alphabetically ordered | Location, mountain range | Length, cm | Altitude, meters a.s.l. |
|---|---|---|---|---|
| 1 | Aghavnadzori Sar 1 | Vardenis | 455 | 1950 |
| 2 | Aghavnadzori Sar 2 | Vardenis | 320 | 1950 |
| 3 | Aghavnadzori Sar 3 | Vardenis | 420 | 2027 |
| 4 | Aghavnadzori Sar 4 | Vardenis | 154 | 1941 |
| 5 | Al Lake 1 | Vardenis | 450 | 2069 |
| 6 | Al Lake 2 | Vardenis | 110 | 2747 |
| 7 | Al Lake 3 | Vardenis | 279 | 2737 |
| 8 | Angeghakot 1 | Syunik | 330 | 1805 |
| 9 | Arshaluysi Sar 1 | Geghama | 221 | 2639 |
| 10 | Arshaluysi Sar 2 | Geghama | 270 | 2645 |
| 11 | Arshaluysi Sar 3 | Geghama | 140 | 2714 |
| 12 | Arshaluysi Sar 4 | Geghama | 198 | 2703 |
| 13 | Arshaluysi Sar 5 | Geghama | 245 | 2670 |
| 14 | Artanish 1 | Areguni | 230 | 1965 |
| 15 | Artanish 2 | Areguni | 290 | 1967 |
| 16 | Artanish 3 | Areguni | 390 | 1922 |
| 17 | Artanish 4 | Areguni | 250 | 1925 |
| 18 | Artanish 5 | Areguni | 220 | 1930 |
| 19 | Aygeshat 1 | Aragats | 140 | 1005 |
| 20 | Aylakh Lake 1 | Vardenis | 330 | 2992 |
| 21 | Bazmaberdi Sar 1 | Aragats | 210 | 1810 |
| 22 | Bnunis 1 | Syunik | 312 | 1875 |
| 23 | Buzhakan 1 | Aragats | 227 | 1818 |
| 24 | Chiva 1 | Vardenis | 170 | 1295 |
| 25 | Darik 1 | Eghnakhagh | 395 | 2147 |
| 26 | Dashtadem 1 | Aragats | 159 | 1438 |
| 27 | Dashtademi Sar 1 | Aragats | 295 | 2336 |
| 28 | Dashtademi Sar 2 | Aragats | 340 | 2313 |
| 29 | Dsoraghbyur 1 | Geghama | 215 | 1576 |
| 30 | Dsoraglukh 1 | Tsaghkunyats | 340 | 2062 |
| 31 | Dsyanberd 1 | Aragats | 290 | 1954 |
| 32 | Eghegisi Sar 1 | Vardenis | 340 | 2264 |
| 33 | Eghnajur 1 | Eghnakhagh | 430 | 2110 |
| 34 | Eghvardi Sar 1 | Aragats | 363 | 2677 |
| 35 | Garni 1 | Geghama | 298 | 1390 |
| 36 | Garni 2 | Geghama | 292 | 1409 |
| 37 | Gaylabulur 1 | Aragats | 162 | 1953 |
| 38 | Geghamasar 1 | Sevan | 360 | 1929 |
| 39 | Gegharda Lake 1 | Geghama | 400 | 2700 |
| 40 | Gegharda Lake 2 | Geghama | 350 | 2700 |
| 41 | Gegharda Lake 3 | Geghama | 195 | 2700 |
| 42 | Gegharda Lake 4 | Geghama | 205 | 2700 |
| 43 | Gegharda Lake 5 | Geghama | 340 | 2700 |
| 44 | Gegharda Lake 6 | Geghama | 315 | 2700 |
| 45 | Gegharda Lake 7 | Geghama | 310 | 2700 |
| 46 | Gegharda Lake 8 | Geghama | 160 | 2700 |
| 47 | Geghasar 1 | Geghama | 390 | 2936 |
| 48 | Geghasar 2 | Geghama | 132 | 2990 |
| 49 | Geghasar 3 | Geghama | 358 | 3174 |
| 50 | Geghashen 1 | Geghama | 193 | 1652 |
| 51 | Glan 1 | Geghama | 480 | 2046 |
| 52 | Goghti Sar 1 | Geghama | 320 | 2141 |
| 53 | Harzhis 1 | Syunik | 184 | 1827 |
| 54 | Harzhis 2 | Syunik | 170 | 1919 |
| 55 | Irind 1 | Aragats | 337 | 1977 |
| 56 | Kakavadsori Sar 1 | Aragats | 385 | 2357 |
| 57 | Karakap 1 | Aragats | 506 | 2770 |
| 58 | Karakap 2 | Aragats | 301 | 2769 |
| 59 | Karakap 3 | Aragats | 359 | 2768 |
| 60 | Karakap 4 | Aragats | 368 | 2769 |
| 61 | Karakap 5 | Aragats | 294 | 2760 |
| 62 | Karmrashen 1 | Vardenis | 366 | 2063 |
| 63 | Khoznavari Sar 1 | Syunik | 238 | 1930 |
| 64 | Koturi Sar 1 | Aragats | 330 | 2297 |
| 65 | Lcharot 1 | Geghama | 329 | 2968 |
| 66 | Lcharot 2 | Geghama | 243 | 2964 |
| 67 | Lcharot 3 | Geghama | 199 | 2963 |
| 68 | Lchashen 1 | Tsaghkunyats | 370 | 1975 |
| 69 | Lchashen 2 | Tsaghkunyats | 265 | 2057 |
| 70 | Lusakn 1 | Aragats | 320 | 1120 |
| 71 | Madinayi Sar 1 | Vardenis | 250 | 2245 |
| 72 | Maghalner 1 | Geghama | 270 | 3013 |
| 73 | Metsadzor 1 | Aragats | 320 | 2086 |
| 74 | Navur 1 | Kenats | 360 | 1422 |
| 75 | Oshakani Sar 1 | Aragats | 332 | 2818 |
| 76 | Sakhurak 1 | Geghama | 550 | 2505 |
| 77 | Sakhurak 2 | Geghama | 375 | 2508 |
| 78 | Sakhurak 3 | Geghama | 330 | 2510 |
| 79 | Sakhurak 4 | Geghama | 260 | 2505 |
| 80 | Sakhurak 5 | Geghama | 293 | 2507 |
| 81 | Sakhurak 6 | Geghama | 330 | 2472 |
| 82 | Sarnaghbyur 1 | Aragats | 360 | 1867 |
| 83 | Sarukhan 1 | Geghama | 305 | 2012 |
| 84 | Sasnasheni Sar 1 | Aragats | 195 | 2020 |
| 85 | Shamirami Sar 1 | Aragats | 325 | 2714 |
| 86 | Sotk 1 | Vardenis | 165 | 2009 |
| 87 | Taratumb 1 | Vardenis | 340 | 1645 |
| 88 | Taratumb 2 | Vardenis | 335 | 1645 |
| 89 | Taratumb 3 | Vardenis | 302 | 1659 |
| 90 | Tirinkatar 1 | Aragats | 295 | 2840 |
| 91 | Tirinkatar 2 | Aragats | 327 | 2849 |
| 92 | Tirinkatar 3 | Aragats | 271 | 2843 |
| 93 | Tirinkatar 4 | Aragats | 426 | 2849 |
| 94 | Tirinkatar 5 | Aragats | 264 | 2851 |
| 95 | Tirinkatar 6 | Aragats | 386 | 2852 |
| 96 | Tirinkatar 7 | Aragats | 287 | 2851 |
| 97 | Tirinkatar 8 | Aragats | 350 | 2846 |
| 98 | Tirinkatar 9 | Aragats | 356 | 2848 |
| 99 | Tirinkatar 10 | Aragats | 285 | 2849 |
| 100 | Tirinkatar 11 | Aragats | 213 | 2863 |
| 101 | Tirinkatar 12 | Aragats | 355 | 2852 |
| 102 | Tsakhkunk 1 | Tsaghkunyats | 375 | 2027 |
| 103 | Ughtuakunk 1 | Geghama | 390 | 2586 |
| 104 | Ughtuakunk 2 | Geghama | 235 | 2587 |
| 105 | Ughtuakunk 3 | Geghama | 295 | 2731 |
| 106 | Vanstan 1 | Geghama | 503 | 2100 |
| 107 | Vanstan 2 | Geghama | 405 | 2256 |
| 108 | Vanstan 3 | Geghama | 390 | 2258 |
| 109 | Vanstan 4 | Geghama | 410 | 2292 |
| 110 | Vanstan 5 | Geghama | 520 | 2338 |
| 111 | Vardenyats 1 | Vardenis | 390 | 2406 |
| 112 | Vishapasar 1 | Geghama | 188 | 2785 |
| 113 | Vishapasar 2 | Geghama | 320 | 2832 |
| 114 | Vosketas 1 | Aragats | 342 | 2065 |
| 115 | Zovunu Sar 1 | Aragats | 340 | 2768 |

A comprehensive geospatial mapping of vishap locations was carried out using GPS and GIS-based spatial analysis. The vishaps' dimensions (height, width, and length) were measured in situ using measuring tapes and laser distance meters to ensure precision. Photogrammetric techniques were employed to create 3D models of selected specimens for comparative morphological analysis. All vishaps were drawn in scale.

Elevation data were recorded using handheld GPS devices and cross-verified with topographic maps and digital elevation models (DEMs).

The correlation between vishap size and altitude was analyzed using statistical regression models. Gaussian distribution models were used to evaluate clustering trends. $R^2$ values were calculated to determine the significance of the observed correlations, cf. [40].

## Results

### Statistical analysis

Utilizing the data presented in Table 1, we conducted a statistical analysis of the correlation between the size (d) of the vishaps and the altitude (h) above sea level (a.s.l.) of their respective locations as reflected in Figure 7a. The distribution of these values is further illustrated in Figure 7b, where a histogram depicts the specific measurements for each object.

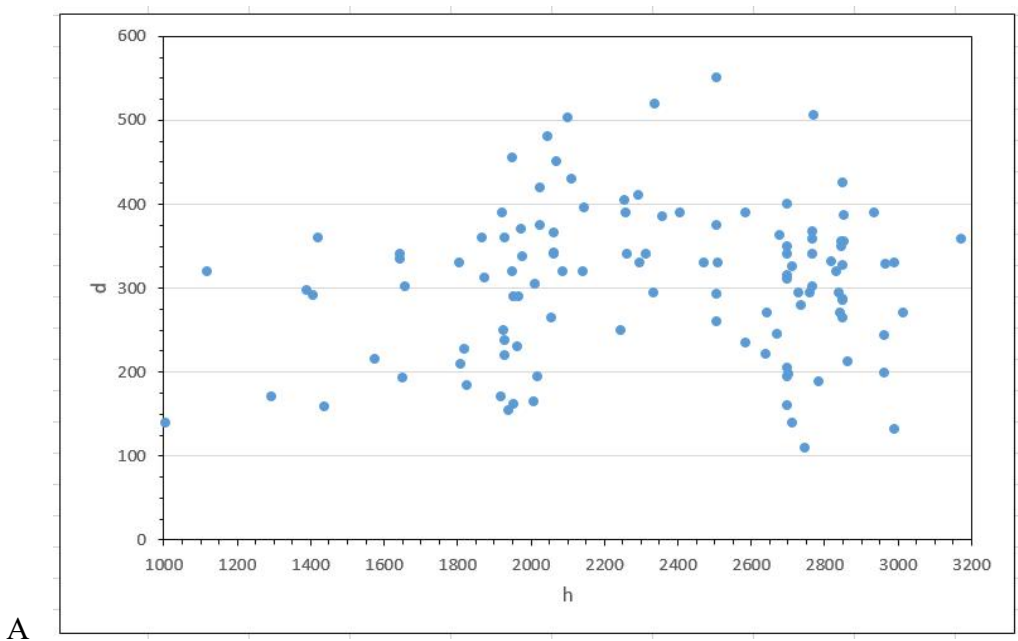

A

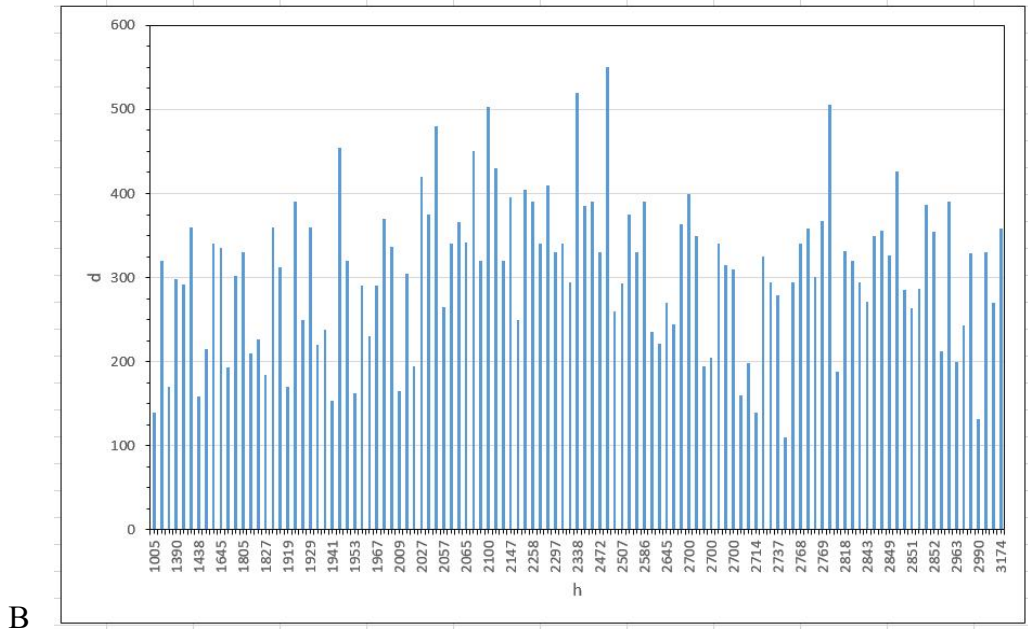

B

*Figure 7. A) The distribution of vishaps' size vs the location altitude above sea level;
B) The same as a) with the altitude values of each object in Table 1 (V. Gurzadyan).*

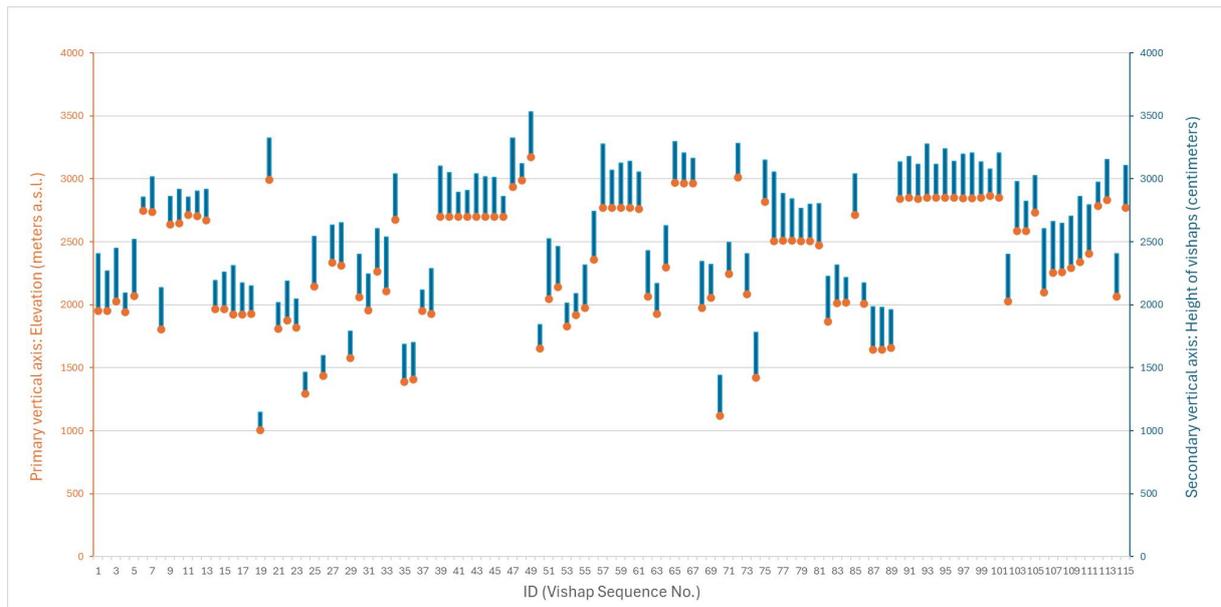

*Figures 8. Correlation of the vishaps length (blue bars) and the altitude of their findspots (orange dots), based on Table 1.
The length (= original height) of the vishaps is indicated by the size of the blue bars, scaled according to the secondary
vertical axis on the right (V. Gurzadyan and P. Hnila).*

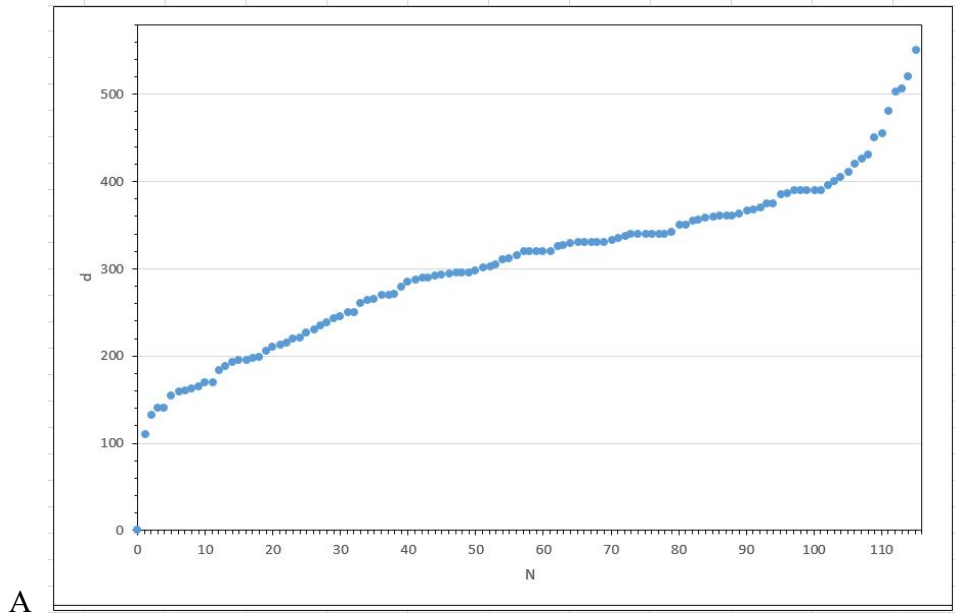

A

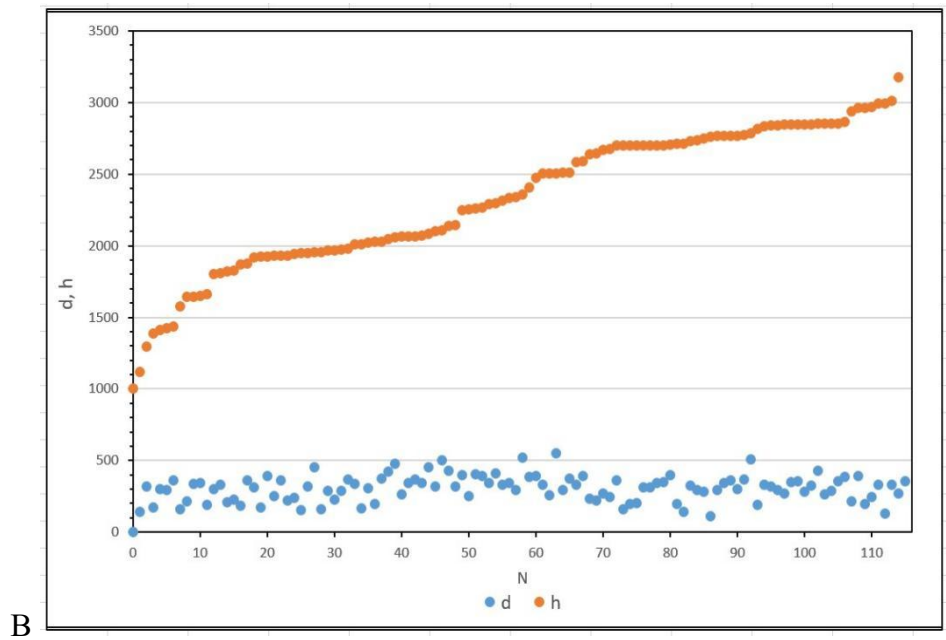

B

*Figure 9 A, B. The ordered distribution of size, d, and altitude, h (V. Gurzadyan).*

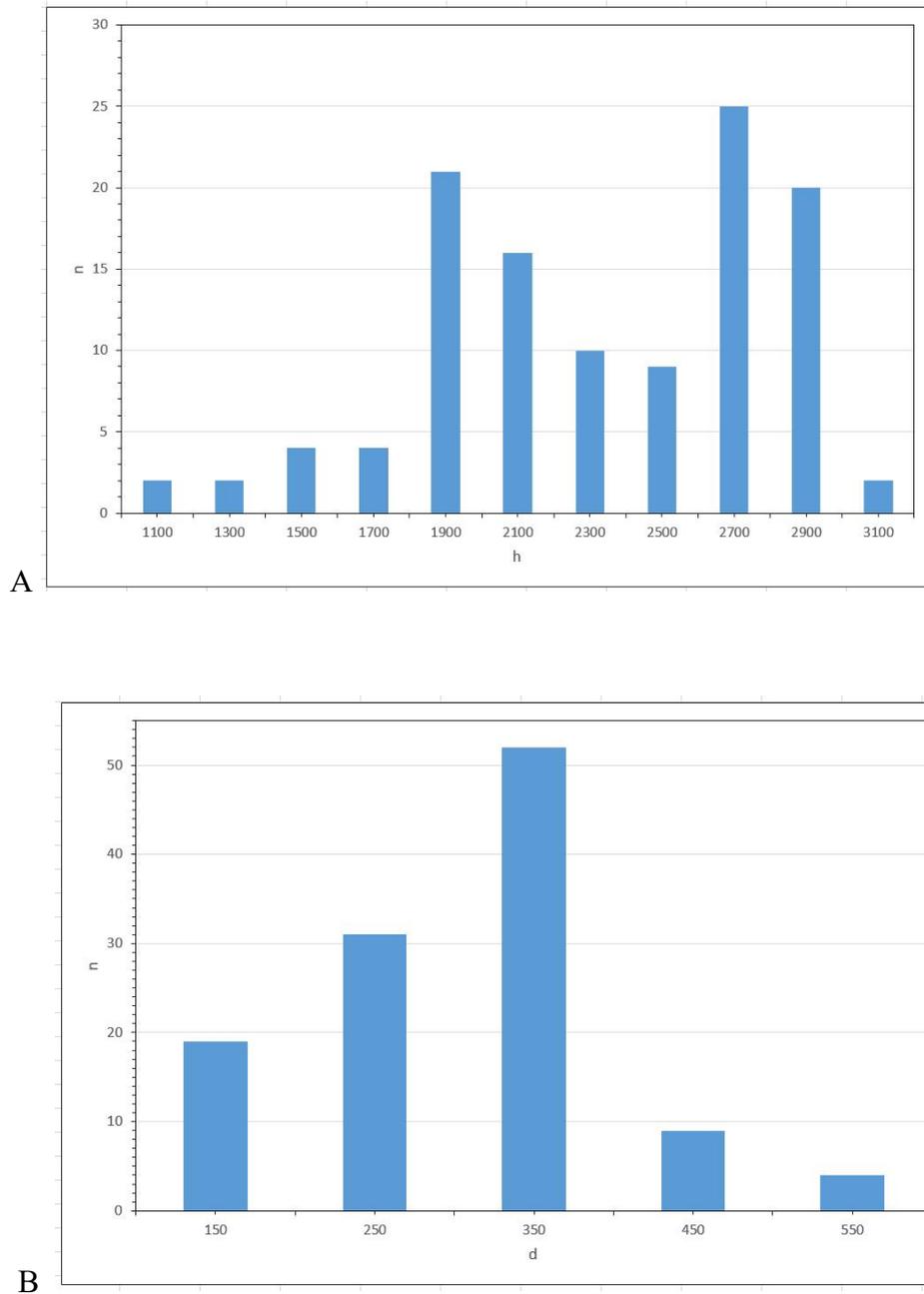

*Figure 10. A) The histogram of the distribution of the altitude of the vishaps; the Gaussian goodness of fit $R^2$ values are, for the left peak, 0.7455 and for the right peak, 0.9885; B) The histogram of the distribution of the sizes of the vishaps: Gaussian goodness of fit $R^2$ value is 0.8383 (V. Gurzadyan).*

Figures 8 and 9 illustrate the distribution of both parameters, while Figure 10 presents histograms depicting the distributions of vishap location altitudes and their corresponding sizes, respectively. The $R^2$ values, representing the goodness of fit for the Gaussian distributions (coefficient of determination) (e.g., [40]), are provided for the two peaks shown in Figure 10a and the single peak in Figure 10b.

The statistical analysis of the physical characteristics of vishaps – specifically their size and altitude – may provide insights into the underlying purposes behind their construction. First, it is important to note that the snow line, which represents the lower topographic boundary of perennial snow cover, plays a significant role in determining the intensity of human activity in mountainous regions, particularly in the Armenian Highlands. For example, Armenian medieval churches are predominantly located in valleys, with a noticeable scarcity of such structures at elevations approaching 2000 meters. Conversely, less labor-intensive constructions, such as megalithic structures, are more commonly found at higher elevations.

The amount of the labor attributed to a given vishap stela typically can be split as follows: (a) the quarrying of a raw piece from the rocks, (b) the processing of the raw stone into a finished, geometrically shaped, and polished monument, occasionally adorned with ornaments, using the technologies available during the prehistoric period, c) transportation of the ready stelae to the final destination. As mentioned above, it is noted that the original location of vishaps correlates with the critical points in prehistoric irrigation systems (Kalantar[6,7,8]), implying transportation of up to 6-7 ton stones to certain fixed areas. The labor associated both to the processing and the transportation of the items is proportional to their size $L$. Namely, the labor for processing (polishing) of a stone depends on its area $S$, which can be approximated as $S(L) \approx 4\pi \{(La)^{1.6} + (Lb)^{1.6}]/3\}^{0.6}$, where $a, b$ are the semi-axes of the elliptic cross section. The transportation labor depends mainly on the weight $W$ of the stelae, also proportional to their length (size): $W(L) = F(L) L \rho$, where $F(L)$ is the cross-section, $\rho$ is the density (e.g. of basalt). Thus, the size of the stelae is a quantitative indicator of the associated labor amount, both for processing and transportation.

Thus, larger vishaps would necessitate greater processing time, especially in regions where the duration of the snow-free period decreases with increasing altitude. Therefore, it might be expected that, at higher elevations, smaller vishaps would be found, assuming that their size and location were not of particular significance to their constructors. It is assumed here that the availability of stone resources is independent of altitude, as stone is widely available at most locations where vishaps have been discovered. Consequently, a statistical correlation between vishap size (d) and altitude (h) could be anticipated. However, the results of our analysis contradicted this hypothesis.

The distribution of vishap sizes, as illustrated in Figure 10, reveals that the mean size peaks at approximately 3.5 meters, with a range extending from smaller to larger specimens, reaching up to over 5.5 meters. However, as shown in Figures 6–10, there is no discernible trend indicating a decrease in the number of larger vishaps with increasing altitude. This suggests that the constructors intentionally devoted their limited activity periods at higher elevations to the construction and transportation of large, labor-intensive monuments, some of which exceed the size of those located at lower altitudes. This is further supported by the continuity of the size-ordered distribution curve depicted in Figure 8a,b.

Thus, the analysis indicates that there were clear motivations to construct vishaps of varying sizes, including large ones, at higher altitudes, despite the additional labor required for their creation and transport in conditions with a much shorter active period. To illustrate the local microclimate vs the activity conditions, one can refer to widely available data of the known ski resort Tsakhkadzor, 1800 m asl (40.54° N 44.71° E): snow available during October-May, maximum mean temperature in October and May, +5°C (as monitored during 2007-2024). At higher elevations, the challenges of supplying food and fuel for workers are significantly greater than at lower altitudes. It is also important to note that even relatively short transportation distances at high altitudes would require substantial human resources, given the weight of the vishaps. For example, the vishap Karakap 3, made of basalt and located at an altitude of approximately 2800 meters, measures 359 x 116 x 44 cm and weighs around 4.3 tons.

Interestingly, as shown in Figure 9, the number of vishaps does not decrease with altitude as expected. Instead, two peaks in the distribution of vishaps are observed at approximately 1900 meters and 2700 meters (Figure 10). These altitudes are associated with drastically different snow levels and, therefore, varying activity periods in the Armenian Highlands. This pattern suggests the existence of specific motivations for constructing and situating vishaps at these higher elevations. These monuments are not concentrated in a single region but are dispersed across broad mountainous areas. While only a few sites contain multiple vishaps at these altitudes, a consistent distribution trend emerges.

The statistical analysis of vishap physical parameters, specifically their size and altitude, indicates that there were deliberate motivations for investing significant labor in constructing these monuments at high altitudes. The natural motivation for locating vishaps at higher elevations may be linked to a cult of water as a life-sustaining force in the valleys below. It is logical to position these objects near summits, which serve as the primary repositories of snow; through continuous melting, it sustains the local populations throughout the year, particularly during the hot and dry summer months.

The clustering of vishaps at distinct altitudes may correlate with seasonal migration patterns or pilgrimages, or both. Notably, both Mount Aragats and the Geghama Mountains serve as significant snow depositories, with abundant water sources such as rivers and lakes. These areas also feature a wide range of archaeological monuments, from megalithic structures to medieval churches and fortresses[7]. The development of an ancient irrigation system in these regions allowed for the efficient distribution of water resources throughout the year[8]. In contrast, Mount Ararat, located directly in front of Mount Aragats, presents a markedly different environment, characterized by porous soils and a limited availability of water sources. This environmental context may account for the relative scarcity of cultural artifacts on its slopes [41-43].

# Discussion

The study of vishap stelae in Armenia, based on their dimensions and altitudinal distribution, provides compelling evidence for their deliberate placement and labor-intensive construction. We adopt the concept of the labor as an informative descriptor of the stelae, namely, the vishap size indicates the amount of the labor for its creation, while their location altitude refers to the limited time (limited labor) for their creation. Then, the findings indicate a general correlation between vishap size and altitude, thus challenging assumptions that larger monuments would be concentrated at lower altitudes. Instead, their presence at high elevations suggests significant cultural motivations, likely tied to the ancient water cult, as vishaps are predominantly located near springs as well as are represented by fish forms. Recall, that human history reveals that usually the cults are indeed associated to significant efforts (labor) of their societies.

The observed bimodal distribution of vishap altitudes at approximately 1900 m and 2700 m suggests a structured pattern of placement. This pattern may be linked to both practical and symbolic importance of high-altitude locations, which is supported by the possible connection of vishaps with ancient irrigation systems, further supporting their functional and cult significance in early societies.

These findings enhance our understanding of high-altitude archaeological sites and the social structures that shaped prehistoric communities. In these regards, comparative studies of vishaps and analogous high-altitude sacred landscapes worldwide (cf., e.g.[44,45]) provide new interpretative perspectives.

**Acknowledgments**

We are thankful to the referees for their valuable comments. We express our sincere gratitude to Prof. Alessandra Gilibert of Ca'Foscari University and to Dr. Pavol Hnila of Free University of Berlin for their valuable contributions during the writing of this paper. We are thankful also to Prof. A. Smith of Cornell University, Dr. K. Bayramyan of the Ministry of Education, Science, Culture, and Sport of Armenia for insightful discussions. Our gratitude extends to the Higher Education and Science Committee of the Ministry of Education, Science, Culture, and Sport of Armenia (grant no. 21AG-6A080) for their co-financing of the "Vishap" Project.

**Data availability**

All relevant data are contained within the manuscript and archaeological materials are stored at the Institute of Archaeology and Ethnography, National Academy of Sciences of Armenia. The site Tirinkatar is part of Armenia's cultural patrimony and is included in UNESCO World Heritage Tentative List.